\documentclass[reprint,superscriptaddress,nofootinbib,amsmath,amssymb,aps,prl,onecolumn]{revtex4-1}

\usepackage[a4paper, margin=2.50cm]{geometry}

\usepackage{bbm}
\usepackage{graphicx}
\usepackage{fancyhdr,color}
\usepackage{hyperref}

\usepackage[T1]{fontenc}
\usepackage[latin9]{inputenc}
\usepackage{mathtools}
\usepackage{amsmath}
\usepackage{stackrel}

\newcommand{\id}{\ensuremath{\mathbbm{1}}}
\newcommand{\rhomic}{\rho_{\mathrm{mc}}}
\newcommand{\rhogmc}{\rho_{\mathrm{gmc}}}

\newcommand{\hr}{{\cal H}}
\newcommand{\tr}{\mbox{Tr}}
\newcommand{\da}{\Delta_{\! A}}
\newcommand{\db}{\Delta_{\! B}}
\newcommand{\dda}{\delta \! A}

\begin{document}

\fancyhead[C]{\sc \color[rgb]{0.4,0.2,0.9}{Quantum Thermodynamics book}}
\fancyhead[R]{}

\title{Dynamical typicality for initial states with 
a preset measurement statistics of
several commuting observables}

\author{Ben N. Balz}
\address{Fakult\"at f\"ur Physik, 
Universit\"at Bielefeld, 
33615 Bielefeld, Germany}

\author{Jonas Richter}
\address{Department of Physics, 
University of Osnabr\"uck, 
49069 Osnabr\"uck, Germany}

\author{Jochen Gemmer}
\address{Department of Physics, 
University of Osnabr\"uck, 
49069 Osnabr\"uck, Germany}

\author{Robin Steinigeweg}
\address{Department of Physics, 
University of Osnabr\"uck, 
49069 Osnabr\"uck, Germany}

\author{Peter Reimann}
\address{Fakult\"at f\"ur Physik, 
Universit\"at Bielefeld, 
33615 Bielefeld, Germany}

\begin{abstract}
We consider all pure or mixed states of a 
quantum many-body system which exhibit the 
same, arbitrary but fixed measurement outcome 
statistics for several commuting observables.
Taking those states as initial conditions, 
which are then propagated by the 
pertinent Schr\"odinger or von Neumann 
equation up to some later time point,
and invoking a few additional, 
fairly weak and realistic 
assumptions, we show that most of them
still entail very similar expectation 
values for any given observable.
This so-called dynamical typicality property
thus corroborates the widespread 
observation that a few macroscopic 
features are sufficient to ensure the 
reproducibility of experimental 
measurements despite many unknown 
and uncontrollable microscopic 
details of the system.
We also discuss and exemplify the 
usefulness of our general analytical 
result as a powerful numerical tool.
\end{abstract}

\maketitle

\thispagestyle{fancy}

\section{Introduction and Overview}
Why are macroscopic experiments reproducible, although 
the molecular details in each repetition of the
experiment are largely unknown and not reproducible\,?
A first path breaking step towards answering this 
question was established by Bartsch and Gemmer \cite{bar09},
putting forward the following so-called dynamical 
typicality property:
The overwhelming majority of all pure states, which exhibit 
very similar expectation 
values of a generic observable at some initial time,
will yield very similar expectation values for this 
observable also at any later time point,
provided the relevant Hilbert space is of large 
(but finite) dimension.
In proving this statement, a technical assumption
is needed, which can strictly speaking only be taken 
for granted if the given initial expectation 
value differs sufficiently little from 
the expectation value in the corresponding
microcanonical equilibrium state
\cite{bar09}.

Here, we work out a significant extension of the 
original dynamical typicality scenario from 
Ref. \cite{bar09}.
Namely, we consider the set of all initial states,
which now may be either pure or mixed, 
and for which all possible measurement 
outcomes for several commuting observables
exhibit certain preset expectation values
(projection probabilities).
Similarly as in \cite{bar09}, we then show
that upon evolving those initial states 
according to the pertinent Schr\"odinger 
or von Neumann equation up to some later 
time point, the vast majority of them still 
gives rise to very similar expectation values 
for any given observable.
But unlike in \cite{bar09}, 
the latter observable may now be 
different from those which determine the 
initial conditions, and the dynamics
may be governed by an arbitrary, 
possibly even explicitly time dependent 
Hamiltonian.

With respect to our analytical explorations, 
certain partial aspects are also related to 
numerous previous works, for instance Refs.
\cite{key-2,key-3,rei07,alv08,sug12,rei18}.
Yet the main physical conclusions
as well as the technical methods 
are quite different.
Concerning the numerical applications,
the basic concepts together with the 
relevant previous literature will be 
reviewed in the last Section.

Our present topic of dynamical typicality
also exhibits certain similarities with
non-dynamical typicality phenomena,
originally due to Lloyd \cite{llo88},
and independently rediscovered 
under the labels ``canonical typicality'' 
and ``concentration of measure effects'' 
in Refs. \cite{gol06,pop06}, 
see also 
\cite{gem04,llo06,rei07,sug07,bal08,tas16}.
In those non-dynamical typicality
explorations, the focus is
on {\em all} quantum states of a
high dimensional Hilbert space,
without any further restriction 
regarding the expectation value 
of some observable, and without
considering the temporal evolution
of those states.
The key result in this context is that
the expectation value of any
given observable is for the overwhelming
majority of those states extremely close
to the pertinent microcanonical (thermal equilibrium)
expectation value of the high dimensional 
Hilbert space under consideration.
A first main virtue of the present
dynamical typicality approach is to 
pick out a small subsets of states with 
distinct non-thermal
features, and then to prove typicality 
properties of their further 
evolution in time.

Another main virtue of such an approach
is to admit statements about the time
evolution without actually solving
the dynamics.
As exemplified by Chaps. 18 and 19,
the latter is a very challenging task in itself
and is outside the conceptual framework
of our present Chapter.
On the other hand, our approach goes beyond 
the scope of Chaps. 18, 19, and 20
in that our systems are in general
not even required to exhibit equilibration 
or thermalization in the long time limit.
Moreover, our dynamical typicality based numerical
scheme can deal with considerably larger systems
than many other methods, such as 
exact diagonalization.

\section{General Framework}
\label{sec:General-Framework}
We consider quantum mechanical model systems, which 
can be described by a Hilbert space $\hr$ of large 
but finite dimension $D$.
The initial state of the system will be specified
in terms of a set of commuting 
observables\footnote{The set of commuting 
observables is not required to be complete,
and even a single observable is admissible.},
whose common eigenspaces are denoted as
$\hr_n$ with $n=1,...,N$.
Hence the projectors $P_n$ onto 
those subspaces $\hr_n$ satisfy 
$P_m P_n=\delta_{mn}P_n$
and $\sum_{n=1}^N P_n=\id_\hr$
(identity on $\hr$),
and the dimensions of the $\hr_n$
are given by
\begin{eqnarray}
d_n =
\tr\{P_n\}
\ ,
\label{1}
\end{eqnarray}
with $D=\sum_{n=1}^Nd_n$.

Denoting by $\rho(0)$ any density operator
(pure or mixed system state)
at the initial time $t=0$, and considering 
the $P_n$'s as observables, the corresponding 
expectation values (projection probabilities) 
are given by
\begin{eqnarray}
p_n= \tr\{\rho(0)\, P_n\} 
\label{2}
\end{eqnarray}
with $p_n\geq 0$ and $\sum_{n=1}^N p_n=1$.
At the focus of our present investigation
will be the set of all initial states 
$\rho(0)$ which satisfy the $N$ constraints (\ref{2})
for arbitrary but fixed values of the 
projection probabilities $\{p_n\}_{n=1}^N$.

Physically speaking, we have mainly quantum 
many-body systems in mind where, 
following von Neumann \cite{key-4}, 
it seems reasonable to expect that
a simultaneous measurement of two or more (almost) 
commuting observables is indeed feasible.
In many cases, some or all of those observables 
will correspond to macroscopic (coarse grained)
quantities and one of them will usually be the 
energy (coarse grained Hamiltonian).
Although such an (approximate) commutation
of those coarse grained observables is usually rather difficult 
to justify more rigorously, it is commonly 
considered as a plausible working hypothesis
\cite{gol10,gol15,tas16,dav85,lin97,has09,oga13}.
For example, we may be dealing with multiple macroscopic 
observables, that can be measured simultaneously, 
or with a single microscopic observable
together with the coarse grained Hamiltonian:
Indeed, it appears as a decent working hypothesis
to assume that the measurement, e.g., of a single particle
velocity leaves the (coarse grained)
energy distribution of
the many-body system practically unaffected. 
On the other hand multiple microscopic observables 
are not expected to commute even approximately. 
It should be emphasized that within this mindest
not the exact Hamiltonian, which governs the dynamics, 
but only its coarse grained counterpart is
imagined to (approximately) commute with 
the coarse grained macroscopic observables.
In particular, the observables are not
required to be conserved quantities.
We also emphasize that this physical interpretation
of the considered mathematical setup will never
be actually used in our subsequent calculations.

Our next observation is that
any real measurement device can only exhibit
a limited number of different possible 
outcomes.
For instance, a digital instrument with $K$ 
digits can only display $10^K$ different
measurement values.
Hence, we can and will assume that 
the number $N$ of common eigenspaces
of all the commuting observables at hand
respects some reasonable bound, say
\begin{equation}
N \leq 10^{20}
 \ .
\label{3}
\end{equation}
Differently speaking, the number of 
commuting observables as well as 
their resolution limits are
assumed to remain experimentally 
realistic.
In principle, also very low $N$ 
values are conceivable and admitted 
in what follows.
In the extreme case, there may be 
just one observable with two 
different possible measurement 
outcomes, implying $N=2$.

On the other hand, for generic macroscopic systems 
with typically $f\approx 10^{23}$ degrees of 
freedom, the dimension $D$ of the relevant Hilbert 
space is exponentially large in $f$.
For instance, $\hr$ may be an energy shell, 
spanned by the eigenvalues of the system 
Hamiltonian with eigenenergies within an 
energy interval,
which is macroscopically small 
(well defined macroscopic system energy) 
but microscopically large 
(exponentially large $D$).
As a consequence, at least one of the
subspaces $\hr_n$ must be extremely 
high dimensional.
More generally, it appears reasonable to assume
that many or even all the $\hr_n$ will exhibit 
a very large dimensionality $d_n$, 
at least as long as 
peculiarities such as experimentally resolvable
gaps in the measurement spectra are absent.
Note that even if all $d_n$'s are large, 
some of them may still be very much 
larger than others.

Compared to the above mentioned
previous works (see Introduction), 
we thus admit the possibility of a
relatively detailed knowledge 
about the initial system state: 
Not only one single expectation value, 
but rather the full statistics of 
all possible measurement outcomes 
of all the commuting observables
is considered to be (approximately) 
fixed via the preset values 
of the projection probabilities $p_n$
in (\ref{2}).
Yet, this information is obviously still 
far from being sufficient to uniquely 
determine the actual microscopic initial
state $\rho(0)$ of the system.

The time evolution of any given initial 
state $\rho(0)$ in (\ref{2}) is governed
by the usual Schr\"odinger or von Neumann
equation.
For our present purposes it is particularly convenient 
to adopt the Heisenberg picture of quantum mechanics
to express the time-dependent expectation value
of any given observable, described by some Hermitian 
operator $A$, as
\begin{eqnarray}
\langle A\rangle _{\!\rho(t)}
& := &
\tr\{ \rho(0)A_t\}
\ , 
\label{4}
\\
A_t 
& := & 
\mathcal{U}_{t}^{\dagger}A\,\mathcal{U}_{t}
\ , 
\label{5}
\end{eqnarray}
where $\mathcal{U}_{t}$ is the quantum-mechanical 
time evolution operator.
For a time-independent Hamiltonian $H$, 
the propagator $\mathcal{U}_{t}$ takes the simple form 
$\exp\{-iHt/\hbar\}$, 
but in full generality, also any explicitly 
time dependent Hamiltonian
$H(t)$ is admitted in (\ref{4}), (\ref{5}).
In particular, the system is not required to 
exhibit equilibration or thermalization
in the long time limit (see also Chap. 18).

Finally, largely arbitrary observables 
$A$ are admitted in (\ref{5}),
except for the weak and physically reasonable 
restriction that the measurement range 
\begin{equation}
\da := a_{\max}-a_{\min}
\ ,
\label{6}
\end{equation}
i.e., the difference between the largest and 
smallest possible eigenvalues 
(or measurement outcomes) of $A$
must be finite, and that the maximal 
resolution $\dda$ of the considered 
measurement device must not be 
unrealistically small compared to $\da$.
For instance, if the measurement
device yields at most $K$ relevant 
digits, then we know that $\da/\dda\leq 10^K$.
Alternatively, $\dda$ may also account
for the finite precision 
when solving the considered quantum 
system by numerical means, 
or simply for the accuracy
with which we actually want 
or need to know the expectation 
value in (\ref{4}).
In any case, the natural reference 
scale for $\dda$ is the measurement 
range from (\ref{6}), i.e., the
appropriate quantity to consider
is the ratio between resolution 
and range,
\begin{eqnarray}
R:=\dda/\da \ .
\label{6a}
\end{eqnarray}

\section{Main Idea and Result}
\label{sec:Main-Result}
In essence, our main idea will be as follows:
We start by distributing all 
initial states $\rho(0)$ compatible 
with (\ref{2}) into different subsets.
Next, we show that the vast majority of 
all $\rho(0)$'s within any given
subset exhibits very similar expectation 
values in (\ref{4}) for an arbitrary 
but fixed $A_t$.
Finally, we will see that the average of 
the expectation values in (\ref{4}) over
all $\rho(0)$'s from one subset
is actually equal for all subsets.
As a consequence, the expectation values
in (\ref{4}) must be very similar for
the vast majority of {\em all} $\rho(0)$'s
which satisfy (\ref{2}), independently 
of the subset to which each of them 
belongs.
In order to show the similarity of (\ref{4}) 
for most $\rho(0)$'s within one subset,
certain (quite weak) assumptions will
be required. 
Once again, these requirements
will turn out to be the same for all 
subsets and thus for all $\rho(0)$ 
which satisfy (\ref{2}).

To begin with, we denote by $U_n$ 
any unitary transformation within 
the subspace $\hr_n$ introduced 
above Eq. (\ref{1}),
i.e., $U_n :\hr_n\to\hr_n$ 
and $U_n^\dagger U_n=\id_{\hr_n}$
(identity on $\hr_n$).
As usual, this operator on $\hr_n$
can be readily ``lifted'' to the
full space $\hr$,
i.e., the same symbol $U_n$ 
now denotes an operator on 
$\hr$ with the key properties 
that $U_n^\dagger U_n=P_n$ and
$U_nP_m=P_m U_n=\delta_{mn} U_n$ 
for all $m,n\in\{1,...,N\}$.
One thus can infer that
\begin{eqnarray}
U:=\sum_{n=1}^N U_n
\label{7}
\end{eqnarray}
is a unitary transformation on $\hr$,
i.e., $U^\dagger U=\id_{\hr}$.
The set of unitaries $U$ which can be
generated via all possible choices
the $U_n$'s in (\ref{7})
is denoted as $S_U$.
One readily confirms that any $U\in S_U$ 
commutes with all the $P_n$'s.
Furthermore, if $\rho(0)$ satisfies
the $N$ constraints (\ref{2}), then 
also 
\begin{eqnarray}
\rho_U(0):=U^\dagger\rho(0)U
\label{7a}
\end{eqnarray}
will do so for all $U\in S_U$.

Any given $\rho(0)$ which satisfies
(\ref{2}) thus generates one of the 
above announced subsets, namely 
$S_{\!\rho(0)}:=\{\rho_U(0)\, | \, U \in S_U\}$.
Obviously, many different $\rho(0)$'s 
which satisfy (\ref{2}) generate 
identical subsets $S_{\!\rho(0)}$.
On the other hand, the union of all 
subsets contains all $\rho(0)$'s 
which satisfy (\ref{2}).
As an aside we note that all
$\rho(0)$'s from the same subset 
exhibit the same spectrum,
but not all $\rho(0)$'s with same 
spectrum belong to the same 
subset.

Finally, we assign a probability
to the $U\in S_U$ 
as follows:
For any given $n$, the $U_n$'s are 
assumed to be uniformly distributed 
(Haar measure with respect to the 
subspace $\hr_n$),
i.e., all of them are equally probable
and statistically independent of each 
other for different indices $n$.
Accordingly, the probability of $U$ in
(\ref{7}) is defined as the joint probability
of all the $U_n$'s appearing on the right hand side,
i.e., each combination of $U_n$'s in (\ref{7}) 
is realized with equal probability.
Averaging any $U$-dependent quantity
$X(U)$ over all $U$'s according
to this probability measure is henceforth
indicated by the symbol $[X(U)]_U$.

Intuitively, this choice is extremely natural.
Indeed, if it is understood that the probability
of any $\rho_U(0)$ within a given
set $S_{\!\rho(0)}$ equals the 
probability of $U$, then our above choice 
is the only one which is consistent, 
i.e., each $\rho(0)$ which generates the same 
set $S_{\!\rho(0)}$ also generates the same 
probability measure on it.

Next we consider $A$ and $t$ in (\ref{5}) and 
thus $A_t$ in (\ref{4}) as arbitrary but fixed.
In general, different $\rho(0)$'s which satisfy
(\ref{2}) are expected to entail different 
expectation values in (\ref{4}).
Focusing on all $\rho(0)$'s which belong
to the same, arbitrary but fixed subset 
$S_{\!\rho(0)}$, one finds for
the average and the variance 
of the expectations values in (\ref{4})
the following results
\begin{eqnarray}
\mu_t
& := & 
\left[\tr \{ \rho_U(0) A_t\} \right]_{U}
=
\sum_{n=1}^{N}
\frac{p_n}{d_{n}}\, \tr\{ A_t P_n\}
\ ,
\label{8}
\\
\sigma_t^2
& := &
\left[ 
\left(
\tr\{ \rho_U(0)A_t \} - 
\mu_t
\right)^2
\right]_U
\leq
\lambda\, \da^2
\ ,
\label{9}
\\
\lambda 
& := &
5\, \underset{n}{\max}\left(\frac{p_n}{d_n}\right)
\ .
\label{10}
\end{eqnarray}
These relations (\ref{8})-(\ref{10}) 
represent the main result of our present work.
However, their detailed derivation
is tedious, not very insightful, 
and hence postponed to the Appendix. 

\section{Discussion}
As announced at the beginning of the previous 
section,
the right hand side of (\ref{8}) is independent
of the actual subset $S_{\!\rho(0)}$, over which 
the average on the left hand side is 
performed, and likewise for (\ref{9}).
Hence, the following conclusion, 
which {\em a priori} implicitly assumes 
that the $\rho(0)$'s are randomly
sampled from one single subset 
$S_{\!\rho(0)}$, 
{\em de facto} also remains true 
when randomly sampling $\rho(0)$'s 
from {\em any} of those subsets,
i.e.,  for {\em all} $\rho(0)$'s which 
satisfy (\ref{2}).
Taking into account this fact and (\ref{9}), 
it follows either obviously or by exploiting 
the so-called Chebyshev (or Markov)
inequality that
\begin{eqnarray}
\mbox{Prob}\left(\,\left|
\tr\{\rho(0)A_t\}-\mu_t
\right|>\epsilon\,\right)
\ & \leq & \,
\lambda\, \, (\da/\epsilon)^2
\label{11}
\end{eqnarray}
for any $\epsilon>0$, where
the probability on the left hand side 
is understood 
as the fraction of all $\rho(0)$'s 
compatible with (\ref{2}), for which
$\tr\{\rho(0)A_t\}$ differs by more 
than $\epsilon$ from the mean value
$\mu_t$.
If we choose for $\epsilon$ 
the pertinent experimental, numerical,
or theoretically required
resolution $\dda$ in (\ref{6a}), then
\begin{eqnarray}
\lambda\ll\, R^2
\label{12}
\end{eqnarray}
implies that $\tr\{\rho(0)A_t\}$ can 
be considered as
indistinguishable\footnote{More sophisticated 
distinguishability measures between quantum 
states than expectation values could in
principle be taken into account along 
similar lines as in Ref. \cite{key-8}.}
from $\mu_t$ for the vast majority 
of all $\rho(0)$'s which satisfy 
(\ref{2}).
In view of (\ref{4}) and (\ref{8})
this means that all those  $\rho(0)$'s
satisfy the approximation
\begin{eqnarray}
\langle A\rangle _{\!\rho(t)}
& = & \tr\{\rhogmc A_t\}
\label{13}
\end{eqnarray}
for our purposes practically perfectly 
well, where
\begin{eqnarray}
\rhogmc & := &
\sum_{n=1}^N p_n \rhomic^{(n)}
\ ,
\label{14}
\\
\rhomic^{(n)} & := & \frac{1}{d_n}P_n
\ .
\label{15}
\end{eqnarray}
With (\ref{1}) one readily sees that 
$\rhomic^{(n)}$ in 
(\ref{15}) amounts to a microcanonical 
density operator on the subspace $\hr_n$,
hence $\rhogmc$ in (\ref{14})
may be viewed as a
``generalized microcanonical ensemble'',
properly accounting for the preset 
weight $p_n$ of each subspace 
in (\ref{2}).

According to the definitions (\ref{1})
and (\ref{2}), the ratio $p_n/d_n$ 
may be viewed as the (mean) population
per eigenstate within any given
eigenspace $\hr_n$.
Recalling that $p_n\geq 0$, 
$\sum_{n=1}^N p_n=1$, and that the total
number of eigenstates $D$ is unimaginably 
large, it is obvious that (\ref{12})
with (\ref{6a}) and (\ref{10})
will be satisfied under many 
very common circumstances.
For instance, one often expects 
(see below (\ref{3})) that the dimension
$d_n$ of every eigenspace $\hr_n$ 
is so large that (\ref{12}) is 
automatically fulfilled without any 
further restricition regarding the $p_n$'s
in (\ref{10}).
Moreover, even if certain
subspaces $\hr_n$ should happen to be 
relatively 
low dimensional, it will be sufficient
that their ``weights'' $p_n$ 
are not unreasonably large (compared
to their relatively low dimensions)
in order to still guarantee (\ref{12}).
Note that this argument even admits that
$d_n=1$ for all $n$, though such cases
seem physically unrealistic according
to the considerations below (\ref{3}).

Finally, it is possible to further generalize
our so far results in the following two ways:

First, there may be one exceptional subspace
$\hr_n$ of dimension $d_n=1$, for which
$p_n$ is not restricted at all.
In other words, one $n$ with $d_n=1$
may be disregarded when taking the
maximum on the right hand side of
(\ref{10}).
For instance, this case may be of interest
for a system with a non-degenerate ground 
state, which is energetically separated 
by a gap from the first excited state
and thus may exhibit an exceptionally
large (macroscopic) population $p_n$ 
compared to all the other level 
populations $p_m/d_m$ with $m\not=n$.
The derivation of this generalization
amounts to a straightforward combination
of the approach in Refs. \cite{key-8,rei12} and 
in the Appendix below, and is therefore 
omitted.

Second, not all weights $p_n$ may be known, 
but, say, only those with indices
$n\in\{1,...,N'\}$, where $N'<N$.
Accordingly, there are only $N'$ constraints 
of the form (\ref{2}) with $p_n\geq 0$ 
and $\sum_{n=1}^{N'}p_n\leq 1$.
It follows that the union (direct sum)
of all the remaining subspaces 
$\tilde\hr:=\hr_{N'+1}\oplus
...\oplus \hr _N$ will be populated 
with probability 
$\tilde p:=1-\sum_{n=1}^{N'}p_n$
and that the projector onto this subspace 
$\tilde\hr$
is given by $\tilde P:=P_{N'+1}+...+P_N$.
Altogether the case at hand can thus be reduced to the
original situation with a new ``effective'' N value,
namely $N=N'+1$, complemented by $P_N:=\tilde P$ and
$p_N:=\tilde p$.

Taking for granted that condition
(\ref{12}) is satisfied, it follows
that the expectation values of $A$ at time
$t$ on the left hand side of (\ref{13})
are practically indistinguishable from 
each other for the vast majority of
all initial states $\rho(0)$ which
satisfy the $N$ constraints (\ref{2}),
i.e., we recover our main dynamical 
typicality result announced in 
the first section.

Note that there may still be a small 
set of ``untypical'' $\rho(0)$'s
which satisfy (\ref{2}) but notably
violate the approximation (\ref{13}).
Recalling that $A$ and $t$ in (\ref{5}) 
and thus $A_t$ in (\ref{4}) are still 
considered as arbitrary but fixed 
(see above (\ref{8})), this set
of ``untypical'' $\rho(0)$'s
will usually be different for 
different time points $t$ and/or
different observables $A$.
In this context, it is worth noting 
that the upper bound in (\ref{9})
does not depend on $t$.
Since averaging over $U$ and 
integrating over $t$ are commuting 
operations, we thus can conclude 
from (\ref{9}) that 
\begin{eqnarray}
Q
& := & 
\left[q_U\right]_U
\leq
\lambda\, \da^2
\ ,
\label{16}
\end{eqnarray}
where 
\begin{eqnarray}
q_U 
& := & 
\frac{1}{t_2-t_1}
\int_{t_1}^{t_2} 
\xi_U^2(t) \, dt
\label{17}
\end{eqnarray}
and
\begin{eqnarray}
\xi_U(t)
& := &
\tr\{ \rho_U(0)A_t \} - \mu_t
\label{18}
\end{eqnarray}
for arbitrary $t_2>t_1\geq 0$.
Since $q_U\geq 0$,
one can conclude as before
from (\ref{12}) and (\ref{16})
that the quantity $q_U$
must be very small for the vast
majority of all $\rho(0)$'s.
For any given such $\rho(0)$, 
also the integrand $\xi_U^2(t)$ 
in (\ref{17}) must be very small 
{\em simultaneously} for all
$t\in [t_1,t_2]$, apart from 
a negligible small fraction
of exceptional times $t$'s.
For sufficiently small $\lambda$ in (\ref{12}),
those exceptional $t$'s become 
unobservably rare and can 
be ignored.
We thus can conclude that
for any given time interval and
any given observable $A$,
most $\rho(0)$'s which satisfy 
(\ref{2}) exhibit very similar
expectation values of $A$
over the entire time interval.

Formally, this typical time evolution 
is given by the right 
hand side of (\ref{13}), but its explicit 
quantitative evaluation is usually 
very difficult (see Chaps. 18 and 19)
and goes beyond the scope of 
our present work.
In fact, it is exactly one of the main 
conceptual virtues of such a dynamical
typicality approach that interesting 
predictions can be obtained without
actually solving the dynamics.
In particular, our present finding helps
us to better understand and explain
the well established fact that
a few macroscopic features are 
sufficient to ensure the reproducibility 
of experimental measurements despite 
many unknown and uncontrollable 
microscopic details of the system.

\section*{Typicality as numerical technique}
As already mentioned at the end of the previous
section, the typical time evolution is given 
by the expression for the ensemble average on 
the right hand side of 
(\ref{13}).
The precise quantitative 
evaluation of this expression can be a 
challenging task for a specific observable 
and many-body quantum model and therefore
it very often has to be done 
numerically (see also Chap. 19).
In this context, the left 
hand side of (\ref{13}) turns out to be 
very useful in order to establish a 
powerful numerical technique 
\cite{sug12,sugiura2013,iitaka2003,elsayed2013, steinigeweg2014-1, 
steinigeweg2016, steinigeweg2017, steinigeweg2014-2,Luitz2017}.
This technique employs the fact that, for a 
fixed initial value (\ref{2}), the ensemble 
average is accurately imitated by a 
single pure state.
This section describes such a dynamical typicality
based numerical method and its central advantages.

To start with, it is convenient to discuss existing numerical approaches. 
Within the large variety of different approaches, a straightforward 
and widely used procedure is the direct evaluation of the ensemble average via
\begin{equation}
\text{Tr} \{ \rho_\text{gmc} \, A_t \} = \sum_{\mu, \nu =1}^D \langle \mu | 
\rho_\text{gmc} | \nu \rangle \langle \nu | A | \mu \rangle \, e^{i (E_\nu - 
E_\mu) t/\hbar} \, ,
\end{equation}
where $| \mu \rangle$, $| \nu \rangle$ and $E_\mu$, $E_\nu$ are the eigenstates 
and corresponding eigenenergies of a given many-body Hamiltonian $H$. In 
principle, these eigenstates and eigenenergies can be calculated numerically by 
the exact diagonalization of systems of finite size. However, because 
the Hilbert-space dimension $D$ grows exponentially fast in the number of 
degrees of freedom, exact diagonalization is only feasible for rather small 
system sizes and finite-size effects can be huge. For a Heisenberg spin-$1/2$ 
chain of length $L$, for example, $D = 2^L$ and $L_\text{max} \sim 20$ is the 
maximum length treatable \cite{elsayed2013, steinigeweg2014-1, steinigeweg2016, 
steinigeweg2017}. For a Fermi-Hubbard chain with $L$ sites, as another important 
example, $D = 4^L$ and $L_\text{max} \sim 10$ is even much less. Clearly, if the 
Hamiltonian $H$ and the observable $A$ have common and mutually commuting 
symmetries, such as total magnetization/particle number or translation 
invariance, it is also possible to exploit these symmetries via
\begin{equation}
\text{Tr} \{ \rho_\text{gmc} \, A_t \} = \sum_{m=1}^M \sum_{\mu, \nu =1}^{d_m} 
\langle m, \mu | \rho_\text{gmc} | m, \nu \rangle \langle m, \nu | A | m, \mu 
\rangle \, e^{i (E_{m,\nu} - E_{m,\mu}) t/\hbar} \, .
\end{equation}
However, using these symmetries for the above examples yields a largest 
subspace with the dimension $d_m$ being 
$d_m \approx (L, L/2)/L$ 
for the Heisenberg spin-$1/2$ chain and 
$d_m \approx (2 L, L)/L$ 
for the Fermi-Hubbard chain \cite{sandvik2010}, where the bracket 
$(n, k): = n!/[k!\cdot(n-k)!]$ 
denotes the binomial coefficient. In fact, symmetries are 
already exploited to reach the aforementioned system sizes $L_\text{max}$.

As compared to exact diagonalization (see Chap. 19), 
a dynamical typicality based 
scheme can treat much larger Hilbert spaces and thus 
allows for significant progress in the 
context of real-time dynamics of expectation values. 
This scheme is based on (\ref{13}) 
with a randomly sampled initial state
of the form
\begin{equation}\label{initial_state}
\rho(0) 
 := 
\frac{1}{2 \, \langle \psi | \rho_\text{gmc} | \psi \rangle} 
\,
(\rho_\text{gmc} \, | \psi \rangle \langle \psi | 
+ \text{h.c.}) \, ,
\end{equation}
with the (unnormalized) pure state
\begin{equation}
 | \psi \rangle 
 :=  
\sum_{k = 1}^D (a_k + i b_k) \, | k \rangle\ ,
\label{initial_state_psi}
\end{equation}
where $|k\rangle$ is an arbitrary but fixed 
orthonormal basis, and
where the $a_k$ and $b_k$ are independent, 
normally distributed random variables 
(i.e. Gaussian distributed with zero mean 
and unit variance).
One readily confirms that all statistical properties
of the random ensemble of pure states in (\ref{initial_state})
are independent of the choice of the basis $|k\rangle$
in (\ref{initial_state_psi}).
In particular, this basis needs not to be the
eigenbasis of $H$.
Rather, any numerically convenient basis
will do the job.
In particular, the basis can be but does 
not need  to be adapted to the symmetries 
of the specific system under consideration.

It readily follows that $\rho(0)^\dagger = 
\rho(0)$ and $\text{Tr} \{ \rho(0) \} = 1$. 
Moreover, by means of similar calculations 
as in the section ``Main idea and results''
one finds that
\begin{equation}
\langle \psi | P_n \, \rho_\text{gmc} | \psi \rangle = \frac{p_n}{d_n} \,
\langle \psi | P_n | \psi \rangle \approx 2 \, p_n
\, , \quad 
\langle \psi | \rho_\text{gmc} | \psi \rangle = \sum_{n=1}^N \frac{p_n}{d_n} \, 
\langle \psi | P_n | \psi \rangle \approx 2
\end{equation}
for sufficiently large subspace dimensions $d_n$. 
For the initial state  $\rho(0)$ in (\ref{initial_state}), 
these two equations lead to the expectation 
value $\text{Tr} \{ P_n \, \rho(0) \} \approx p_n$, 
i.e., the condition in 
(\ref{2}). Therefore, (\ref{13}) applies to this initial state and yields
\begin{equation}
\text{Tr} \{ \rho_\text{gmc} \, A_t \} \approx \frac{\langle \psi | \, 
(\rho_\text{gmc} \, A_t + \text{h.c.}) \, | \psi \rangle}{2 \, \langle \psi | 
\, \rho_\text{gmc} \, | \psi \rangle} = \frac{\text{Re} \, \langle \psi | \, 
\rho_\text{gmc} \, A_t \, | \psi \rangle}{\langle \psi | \, \rho_\text{gmc} \, | 
\psi \rangle}
\end{equation}
and, using $\langle \psi | \rho_\text{gmc} | 
\psi \rangle \approx 2$ again, 
as well as $\langle \psi | \psi \rangle \approx  2 \, D$, 
one obtains the relation
\begin{equation}
\text{Tr} \{ \rho_\text{gmc} \, A_t \} \approx D \, \frac{\text{Re} \, \langle 
\psi | \, \rho_\text{gmc} \, A_t \, | \psi \rangle}{\langle \psi | \psi 
\rangle} \, .
\end{equation}
Thus, the trace $\text{Tr} \{ \bullet \}$ is essentially replaced by a scalar 
product $\langle \psi | \bullet | \psi \rangle$ involving a single pure state 
drawn at random from a Hilbert space of finite but high dimension $D$. This 
relation can be also written in the form
\begin{equation}
\text{Tr} \{ \rho_\text{gmc} \, A_t \} \approx D\, \frac{\text{Re} \, \langle 
\Phi(t)| \, A \, | \varphi(t) \rangle}{\langle \varphi(0) | \varphi(0) \rangle} 
\label{relation}
\end{equation}
with the two auxiliary pure states
\begin{equation}
| \varphi(t) \rangle := e^{-i H t/\hbar} \, | \psi \rangle \, , \quad | \Phi(t) 
\rangle := e^{-i H t/\hbar} \, \rho_\text{gmc} \, | \psi \rangle \, . 
\label{states}
\end{equation}
Note that these pure states look similar but differ from each other because of 
the additional operator $\rho_\text{gmc}$ in the definition of $| \Phi(t) 
\rangle$.

A central advantage of the relation in (\ref{relation}) is that no time 
dependence of operators occurs. Instead, the full time dependence appears as a 
property of pure states only. As a consequence, one does no need to (i) employ 
exact diagonalization and (ii) store full matrices in computer memory. To see 
that exact diagonalization can be circumvented, consider the Schr\"odinger 
equation
\begin{equation}
\frac{\text{d}}{\text{d} t} \, | \varphi(t) \rangle = -\frac{i}{\hbar} \, H \, 
| \varphi(t) 
\rangle
\end{equation}
for $| \varphi(t) \rangle$. This differential equation and the corresponding 
equation for $| \Phi(t) \rangle$ can be solved numerically by straightforward 
iterator schemes like fourth-order Runge-Kutta \cite{elsayed2013, 
steinigeweg2014-1, steinigeweg2016} or more sophisticated schemes 
like Trotter decompositions or Chebyshev polynomials \cite{weisse2006, 
deraedt2006}. Still, one has to implement the action of the Hamiltonian on pure 
states. Since it is possible to carry out these matrix-vector multiplications 
without storing matrices in computer memory, the memory requirement of the 
algorithm is set only by the size of vectors, i.e., ${\cal O}(D)$ or, in case 
of symmetry reduction, ${\cal O}(d_m)$. Nevertheless, to reduce the run time of 
the algorithm, it is convenient to store at least parts of matrices in computer 
memory. In this respect, one can profit from the fact that the Hamiltonian is 
usually a few-body operator with a sparse-matrix representation. Thus, the 
memory requirement essentially remains linear in the relevant Hilbert-space 
dimension. In this way, for the above example of a Heisenberg spin-$1/2$ chain, 
$L_\text{max} = 34$ \cite{steinigeweg2014-1} has been reached using 
medium-sized clusters while $L_\text{max} > 34$ is feasible using massive 
parallelization and supercomputers \cite{steinigeweg2017, worldrecord}. As 
compared to exact diagonalization, the corresponding Hilbert space dimension is 
larger by several orders of magnitude, e.g., by a factor $2^{34}/2^{20} = {\cal 
O}(10^4)$. For such finite but already huge Hilbert spaces, the 
typicality-related approximation error has been demonstrated to be negligibly 
small, by a detailed comparison with other state-of-the-art numerical methods 
including time-dependent density-matrix renormalization group 
\cite{steinigeweg2014-1} and Lanczos diagonalization \cite{steinigeweg2016}.

To illustrate the accuracy of the numerical method and the validity of the 
analytical considerations of the last sections in detail, let us consider the 
Heisenberg spin-$1/2$ chain and the specific operators $n_r = S_r^z + 1/2$, 
where $S_r^z$ is the $z$-component of a spin at site $r$. These $L$ local 
operators are mutually commuting and can be adjusted independently between $0$ 
and $1$. We choose initial conditions $\rho(0)$ with $\langle n_r 
\rangle_{\rho(0)} = 1$ at a single site $r$ and $\langle n_{r'} 
\rangle_{\rho(0)} = 1/2$ at all other sites $r' \neq r$. This choice corresponds 
to $\rho_\text{gmc} = n_r/2^{L-1}$ and, for the observable $A = n_r$, the 
ensemble average is then given by the two-point correlation function $\text{Tr} 
\{ \rho_\text{gmc} \, A_t \} = \text{Tr} \{ n_r n_r(t) \}/2^{L-1}$ at formally infinite temperature,
see (\ref{32}) below or Ref. \cite{steinigeweg2017}. 
In Fig.\ 1 (a) we show this 
ensemble average, as obtained from exact diagonalization for a finite lattice
with $L = 14$ sites. We further depict the approximation in Eq.\ 
(\ref{relation}), as obtained from the Runge-Kutta propagation of a random 
pure state. Clearly, both curves are very close to each other, especially in 
view of the small $L = 14$. Moreover, an agreement of the same quality is 
found for another random pure state. In Fig.\ 1 (c) we compare the 
approximation in Eq.\ (\ref{relation}) for two different pure states drawn at 
random and a substantially larger lattice with $L=28$ sites. Apparently, the 
corresponding curves are much closer to each other. This closeness demonstrates 
the fact that the approximation in Eq.\ (\ref{relation}) becomes exact in the 
thermodynamic limit $L \to \infty$.

\begin{figure}[tb]
\includegraphics[width=\columnwidth]{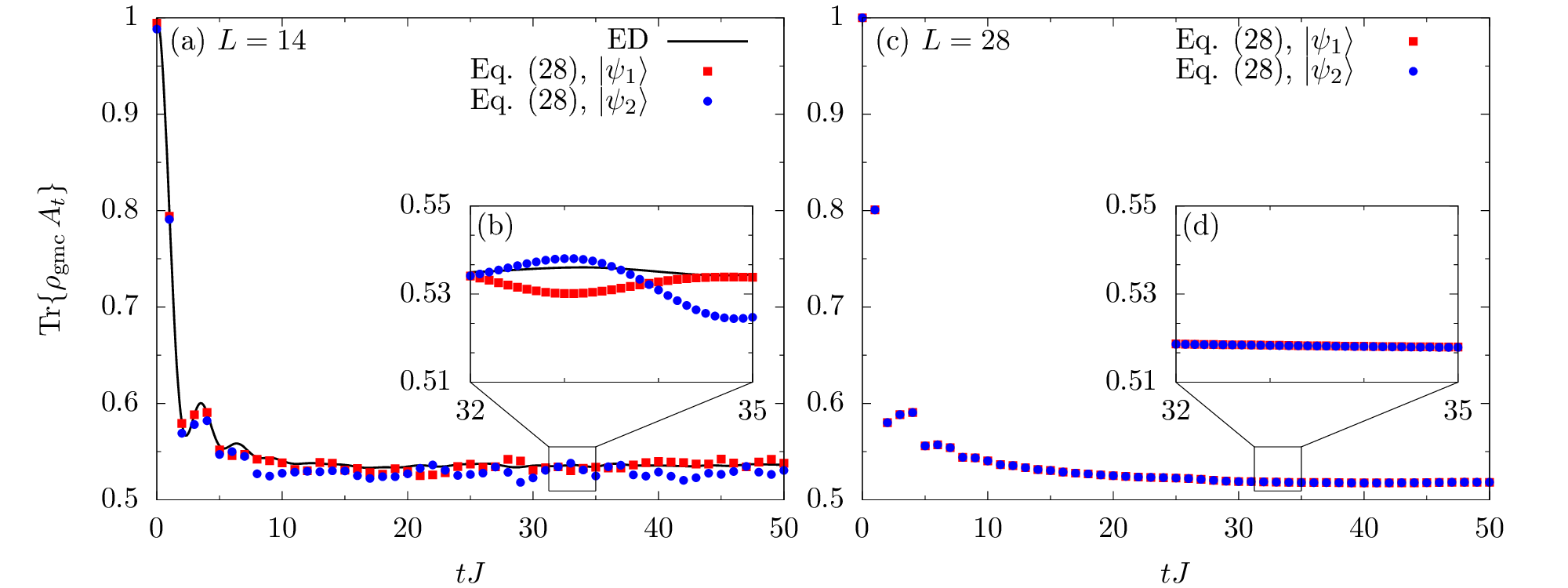}
\caption{(a) Time-dependent expectation value $\text{Tr} \{ \rho_\text{gmc} \, 
A_t \}$ for $\rho_\text{gmc} = n_r$, $A = n_r$, and a Heisenberg spin-$1/2$ 
chain of length $L=14$. Exact diagonalization (ED) is compared to the 
approximation in Eq.\ (\ref{relation}) and two states $| \psi_1 \rangle$ and $| 
\psi_2 \rangle$ drawn at random. (c) Approximation in Eq.\ (\ref{relation}) for 
two random states $| \psi_1 \rangle$ and $| \psi_2 \rangle$ in a lattice with 
$L = 28$ sites. (b), (d) zoom of (a), (c).}
\label{Fig1}
\end{figure}

Remarkably, typicality also provides the basis of a numerical approach to 
autocorrelation functions at finite temperatures. This approach is based on the 
relation \cite{elsayed2013, steinigeweg2014-1, steinigeweg2016}
\begin{equation}
\frac{\text{Re} \, \text{Tr} \{ e^{-\beta H} A A_t \}}{\text{Tr} \{ e^{-\beta 
H} \}} \approx \frac{\text{Re} \, \langle \psi | \, e^{-\beta H} A A_t 
\, | \psi \rangle}{\langle \psi | \, e^{-\beta H} \, | \psi \rangle } \, ,
\label{32}
\end{equation}
where $\beta = 1/k_\text{B} T$ denotes the 
inverse temperature and $| \psi \rangle$ 
is a randomly sampled pure state as defined before. 
Thus, once again, the trace 
$\text{Tr} \{ \bullet \}$ is replaced by the scalar product $\langle \psi | 
\bullet | \psi \rangle$. Similar to (\ref{relation}), this relation can be 
rewritten as
\begin{equation}
\frac{\text{Re} \, \text{Tr} \{ e^{-\beta H} A A_t \}}{\text{Tr} \{ e^{-\beta 
H} \}} \approx \frac{\text{Re} \, \langle \Phi_\beta(t) | \, A \, | 
\varphi_\beta(t) \rangle}{\langle \varphi_\beta(0) | \varphi_\beta(0) \rangle} 
\label{relation2}
\end{equation}
with the two modified auxiliary pure states
\begin{equation}
| \varphi_\beta(t) \rangle := e^{-i H t/\hbar} \, e^{-\beta H/2} \, | \psi 
\rangle \, , \quad | \Phi(t) \rangle := e^{-i H t/\hbar} \, A \, e^{-\beta H/2} 
\,| \psi \rangle \, . \label{states2}
\end{equation}
An important difference between these pure states and the ones in (\ref{states})
is the additional occurrence of the imaginary time $\beta$. As the dependence 
on real time $t$, this dependence can be obtained from iteratively solving  
an imaginary-time Schr\"odinger equation,
\begin{equation}
\frac{\text{d}}{\text{d} \beta} \, | \varphi_\beta(0) \rangle = -\frac{1}{2} \, 
H \, | \varphi_\beta(0) \rangle
\end{equation}
for $| \varphi_\beta(0) \rangle$ and analogously for $| \Phi_\beta(0) \rangle$. 
Certainly, typicality arguments justify the relation in (\ref{relation2}) only, 
if the dimensions of the relevant energy subspaces are sufficiently large. 
If the inverse temperature $\beta$ is small, the dominant contribution stems 
from energy subspaces far from the borders of the spectrum. However, if the 
inverse temperature $\beta$ is large, the dominant contribution stems from 
small energy subspaces close to the lower border of the spectrum and typicality 
arguments cannot be employed any further \cite{steinigeweg2014-2}.

Interestingly, however, this relation does not break down in the 
limit $\beta \to \infty$ and even becomes exact in this limit. This fact 
follows from the definition of the two pure states in (\ref{states2}) and 
whenever the random pure state $| \psi \rangle$ has a finite overlap with the 
ground state $| \mu_0 \rangle$,
\begin{equation}
| \langle \mu_0 | \psi \rangle | > 0 \, ,
\end{equation}
even if this overlap is tiny. Then,
\begin{equation}
\lim_{\beta \to \infty} \frac{| \varphi_\beta(t) \rangle}{\sqrt{\langle 
\varphi_\beta(0) | \varphi_\beta(0) \rangle}} = e^{-i H t/\hbar} \, | \mu_0 
\rangle \, , \quad \lim_{\beta \to \infty} \frac{| \Phi_\beta(t) 
\rangle}{\sqrt{\langle \varphi_\beta(0) | \varphi_\beta(0) \rangle}} = e^{-i H 
t/\hbar} \, A \, | \mu_0 \rangle 
\end{equation}
and, as a consequence,
\begin{equation}
\lim_{\beta \to \infty} \frac{\text{Re} \, \langle \Phi_\beta(t) | \, A \, | 
\varphi_\beta(t) \rangle}{\langle \varphi_\beta(0) | \varphi_\beta(0) \rangle} 
= \text{Re} \, \langle \mu_0 | \, A A_t \, | \mu_0 \rangle \, .
\end{equation}
Thus, in conclusion, Eq. \eqref{relation2} 
also provides a reasonable numerical approach to low temperatures.

\section*{Appendix}
This appendix provides the detailed derivation of the main result
(\ref{8})-(\ref{10}) of our present work.

For notational simplicity, we adopt the abbreviations 
\begin{eqnarray}
\rho & := & \rho(0)\label{19}\\
B & := & A_{t}\ ,\label{20}
\end{eqnarray}
where $A_{t}$ is defined in (\ref{5}). It follows that the eigenvalues
of $B$ are identical to those of $A$. In particular, the measurement
range $\db$ of $B$ will agree with the measurement range $\da$
of $A$ from (\ref{6}), i.e., 
\begin{eqnarray}
\da=\db\ .\label{21}
\end{eqnarray}

\subsection*{Derivation of (\ref{8})}
For any given $n\in\{1,...,N\}$, we denote by $\{|n,a\rangle\}_{a=1}^{d_{n}}$
an arbitrary but fixed orthonormal basis of $\hr_{n}$. The union
of all those bases thus amounts to an orthonormal basis of $\hr$,
i.e., 
\begin{eqnarray}
 &  & \langle n,a|m,b\rangle=\delta_{mn}\delta_{ab}\label{24}\\
 &  & \sum_{n=1}^{N}\sum_{a=1}^{d_{n}}\left|n,a\right\rangle \left\langle n,a\right|=\id_{\hr}\ .\label{25}
\end{eqnarray}
Inserting $1_{\mathcal{H}}$ from (\ref{25}) three times into the
definition of $\mu_{t}$ in (\ref{8}) and observing (\ref{7a}) yields
\begin{eqnarray}
\mu_{t} & =\left[\sum_{l,j,m,n}\sum_{\alpha,\beta,a,b}\left\langle l,\alpha\right|U^{\dagger}\left|m,a\right\rangle \left\langle m,a\right|\rho\left|n,b\right\rangle \left\langle n,b\right|U\left|j,\beta\right\rangle \left\langle j,\beta\right|B\left|l,\alpha\right\rangle \right]_{U}\nonumber \\
 & =\left[\sum_{m,n}\sum_{\alpha,\beta,a,b}\left\langle m,\alpha\right|U^{\dagger}\left|m,a\right\rangle \left\langle m,a\right|\rho\left|n,b\right\rangle \left\langle n,b\right|U\left|n,\beta\right\rangle \left\langle n,\beta\right|B\left|m,\alpha\right\rangle \right]_{U}\,.\label{eq:13}
\end{eqnarray}
In the last step above we used $\left\langle n,b\right|U\left|m,\beta\right\rangle =\delta_{mn}\left\langle n,b\right|U\left|n,\beta\right\rangle $
since any $U\in S_{U}$ commutes with all the $P_{n}$'s as mentioned
below \eqref{7}. Now, we define the abbreviation for the basis representation
of the $U_{n}$ introduced above (\ref{7}) as 
\begin{equation}
U_{n,b\beta}:=\left\langle n,b\right|U_{n}\left|n,\beta\right\rangle =\left\langle n,b\right|U\left|n,\beta\right\rangle .
\end{equation}
This enables us to rewrite equation (\ref{eq:13}) as 
\begin{equation}
\mu_{t}=\sum_{m,n}\sum_{\alpha,\beta,a,b}\left\langle m,a\right|\rho\left|n,b\right\rangle \left\langle n,\beta\right|B\left|m,\alpha\right\rangle \left[U_{m,a\alpha}^{*}U_{n,b\beta}\right]_{U}.\label{eq:15}
\end{equation}

To continue evaluating this expression we revert to \cite{key-9}.
There, the following is stated for the average over uniformly (Haar)
distributed unitaries\footnote{Averages of this type were analyzed by many authors \cite{key-10,key-11,key-12,key-13,key-14,gem04},
often independently. We refer to \cite{key-9}, since only there cases
up to $n\leq5$ are provided, although in the following $n\leq2$.}: 
\begin{equation}
\left[U_{l,a_{1}b_{1}}...U_{l,a_{m}b_{m}}U_{l,\alpha_{1}\beta_{1}}^{*}...U_{l,\alpha_{n}\beta_{n}}^{*}\right]_{U_{l}}=\delta_{mn}\sum_{\varPi,\varPi^{'}}V_{\varPi,\varPi'}\prod_{j=1}^{n}\delta_{a_{j}\alpha_{\varPi(j)}}\delta_{b_{j}\beta_{\varPi'(j)}}.\label{eq:16}
\end{equation}
Quoting verbatim from \cite{key-9}: ``the summation is over all
permutations $\varPi$ and $\varPi'$ of the numbers $1,...,n$. The
coefficients $V_{\varPi,\varPi'}$ depend only on the cycle structure
of the permutation $\varPi^{-1}\varPi'$. Recall that each permutation
of $1,...,n$ has a unique factorization in disjoint cyclic permutations
(``cycles'') of lengths $c_{1},...,c_{k}$ (where $n=\sum_{j=1}^{k}c_{j}$).
The statement that $V_{\varPi,\varPi'}$ depends only on the cycle
structure of $\varPi^{-1}\varPi'$ means that $V_{\varPi,\varPi'}$
depends only of the lengths $c_{1},...,c_{k}$ of the cycles in the
factorization of $\varPi^{-1}\varPi'$. One may therefore write $V_{c_{1},...,c_{k}}$
instead of $V_{\varPi,\varPi'}$.'' The factors $V_{c_{1},...,c_{k}}$
are given by the columns ``CUE'' of Tables II and IV in \cite{key-9}.

As said below (\ref{7a}), the symbol $[...]_{U}$ indicates an average
over all $U_{n}$ within each $\mathcal{H}_{n}$ according to the
Haar measure. Averages over Haar measures in distinct eigenspaces
are statistically independent (see below (\ref{7a})). Thus, by applying
(\ref{eq:16}) to (\ref{eq:15}) we see that $m$ has to equal $n$
or the average vanishes 
\begin{align}
\mu_{t} & =\sum_{n=1}^{N}\sum_{\alpha,\beta,a,b}\left\langle n,a\right|\rho\left|n,b\right\rangle \left\langle n,\beta\right|B\left|n,\alpha\right\rangle \left[U_{n,a\alpha}^{*}U_{n,b\beta}\right]_{U_{n}}\\
 & =\sum_{n=1}^{N}\sum_{\alpha,\beta,a,b}\left\langle n,a\right|\rho\left|n,b\right\rangle \left\langle n,\beta\right|B\left|n,\alpha\right\rangle V_{\varPi_{1},\varPi_{1}}\delta_{a\alpha}\delta_{b\beta}.\label{eq:18}
\end{align}
Here, there is only one permutation: namely the identity, denoted
by $\varPi_{1}$. In this case $\varPi_{1}^{-1}\varPi_{1}=\varPi_{1}$
yields only one cycle of length $c_{1}=1$. The corresponding $V_{\varPi_{1}}=V_{1}$
can be found, in the Tables II and IV in \cite{key-9} (column ``CUE'',
row $n=1$) as $V_{\varPi_{1}}=\frac{1}{d_{n}}$. Plugging this back
into the above equation we obtain 
\begin{equation}
\mu_{t}=\sum_{n=1}^{N}\frac{1}{d_{n}}\tr\left\{ \rho P_{n}\right\} \tr\left\{ BP_{n}\right\} .\label{eq:32}
\end{equation}
Observing (\ref{2}) and (\ref{20}), we finally recover (\ref{8}).

\subsection*{Derivation of (\ref{9}) and (\ref{10})}
The variance in \eqref{9} can be rewritten as 
\begin{equation}
\sigma_{t}^{2}=\left[\tr\left\{ \rho_U B\right\} ^{2}\right]_{U}-\mu_{t}^{2}\,.
\end{equation}
On the right hand side, the last term $\mu_{t}^{2}$ follows from
\eqref{eq:32} above.
Turning to the first term, one finds similarly
as in \eqref{eq:13} and \eqref{eq:15} that
\begin{align}
\left[\tr\left\{ \rho_U B\right\} ^{2}\right]_{U} & =\left[\sum_{l,j}\sum_{\alpha,\beta,a,b}U_{l,a\alpha}^{*}U_{j,b\beta}\left\langle l,a\right|\rho\left|j,b\right\rangle \left\langle j,\beta\right|B\left|l,\alpha\right\rangle \right.\nonumber \\
 & \cdot\left.\sum_{m,n}\sum_{\gamma,\omega,c,d}U_{m,c\gamma}^{*}U_{n,d\omega}\left\langle m,c\right|\rho\left|n,d\right\rangle \left\langle n,\omega\right|B\left|m,\gamma\right\rangle \right]_{U}\,.\label{eq:34-1}
\end{align}
For the summation indices $l,j,m,n$ we define the set of quadruples
\begin{equation}
S\coloneqq\left\{ \left(l,j,m,n\right)\left|l,j,m,n\in\left\{ 1,...,N\right\} \right.\right\} \,.
\end{equation}
If one reverts back to equation (\ref{eq:16}), one finds the necessity
to study three cases for which the average in \eqref{eq:34-1} does
not vanish, which amount to the following index subsets of $S$:~
$S_{1}:\,l=j\neq m=n$, $S_{2}:\,l=j=m=n$ and $S_{3}:\,l=n\neq m=j$.
From now on, the $S_{i}$ refer not only to the corresponding index
sets but to the associated terms in \eqref{eq:34-1}. Thus the following
holds 
\begin{equation}
\left[\tr\left\{ \rho_U B\right\} ^{2}\right]_{U}=\left[S_{1}\right]_{U}+\left[S_{2}\right]_{U}+\left[S_{3}\right]_{U}\,.
\end{equation}

In the first case, $S_{1}$, the average over $U$ factorizes into
two averages over the eigenspaces labeled by $l$ and $m$. Hence,
the calculation reduces to the derivation of (\ref{8}) and we find
\begin{align}
\left[S_{1}\right]_{U} & =\sum_{l\neq m}\sum_{\alpha,\beta,a,b}\left\langle l,a\right|\rho\left|l,b\right\rangle \left\langle l,\beta\right|B\left|l,\alpha\right\rangle \left[U_{l,a\alpha}^{*}U_{l,b\beta}\right]_{U_{l}}\sum_{\gamma,\omega,c,d}\left\langle m,c\right|\rho\left|m,d\right\rangle \left\langle m,\omega\right|B\left|m,\gamma\right\rangle \left[U_{m,c\gamma}^{*}U_{m,d\omega}\right]_{U_{m}}\nonumber \\
 & =\sum_{l\neq m}\frac{1}{d_{l}}\tr\left\{ \rho P_{l}\right\} \tr\left\{ BP_{l}\right\} \frac{1}{d_{m}}\tr\left\{ \rho P_{m}\right\} \tr\left\{ BP_{m}\right\} \label{eq:21}
\end{align}

In the second case, $S_{2}$, the unitary average is not the same
as performed before, but instead: 
\begin{align}
\left[S_{2}\right]_{U} & =\sum_{l=1}^{N}\sum_{\alpha,\beta,a,b}\left\langle l,a\right|\rho\left|l,b\right\rangle \left\langle l,\beta\right|B\left|l,\alpha\right\rangle \sum_{\gamma,\omega,c,d}\left\langle l,c\right|\rho\left|l,d\right\rangle \left\langle l,\omega\right|B\left|l,\gamma\right\rangle \left[U_{l,a\alpha}^{*}U_{l,b\beta}U_{l,c\gamma}^{*}U_{l,d\omega}\right]_{U_{l}}\label{eq:40}
\end{align}
 To further evaluate this expression we make again use of \eqref{eq:16},
arriving at 
\begin{align}
\left[U_{l,b\beta}U_{l,d\omega}U_{l,a\alpha}^{*}U_{l,c\gamma}^{*}\right]_{U_{l}} & =\underset{1/\left(d_{l}^{2}-1\right)}{\underbrace{V_{\varPi_{1},\varPi_{1}}}}\delta_{ba}\delta_{dc}\delta_{\beta\alpha}\delta_{\omega\gamma}+\underset{-1/\left(d_{l}\left(d_{l}^{2}-1\right)\right)}{\underbrace{V_{\varPi_{1},\varPi_{2}}}}\delta_{ba}\delta_{dc}\delta_{\beta\gamma}\delta_{\omega\alpha}\nonumber \\
 & +\underset{-1/\left(d_{l}\left(d_{l}^{2}-1\right)\right)}{\underbrace{V_{\varPi_{2},\varPi_{1}}}}\delta_{bc}\delta_{da}\delta_{\beta\alpha}\delta_{\omega\gamma}+\underset{1/\left(d_{l}^{2}-1\right)}{\underbrace{V_{\varPi_{2},\varPi_{2}}}}\delta_{bc}\delta_{da}\delta_{\beta\gamma}\delta_{\omega\alpha}\,.\label{41}
\end{align}
Here, $\varPi_{1}$ still equates to the identity, but now $\varPi_{2}$
denotes the permutation exchanging the numbers $1$ and $2$. Therefore,
$\varPi_{1}^{-1}\varPi_{1}=\varPi_{2}^{-1}\varPi_{2}=\varPi_{1}$
decomposes in two cycles of lengths $c_{1}=c_{2}=1$, while $\varPi_{2}^{-1}\varPi_{1}=\varPi_{1}^{-1}\varPi_{2}=\varPi_{2}$
exhibits one cycle with $c_{2}=2$. The corresponding $V_{\varPi_{1},\varPi_{1}}=V_{\varPi_{2},\varPi_{2}}=V_{1,1}$
and $V_{\varPi_{2},\varPi_{1}}=V_{\varPi_{1},\varPi_{2}}=V_{2}$ can
again be found, in the Tables II and IV in \cite{key-9} (column ``CUE'',
row $n=2$) with the above cited results. Plugging \eqref{41} back
into \eqref{eq:40} and additionally using $1/\left(d_{l}^{2}-1\right)=1/d_{l}^{2}+1/\left(d_{l}^{2}\left(d_{l}^{2}-1\right)\right)$,
we obtain 
\begin{eqnarray}
\left[S_{2}\right]_{U} & = & \sum_{l=1}^{N}\frac{1}{d_{l}^{2}}\tr\left\{ \rho P_{l}\right\} ^{2}\tr\left\{ BP_{l}\right\} ^{2}+S_{2}^{'}\,,\label{eq:23}
\end{eqnarray}
\begin{align}
S_{2}^{'} & \coloneqq\sum_{l=1}^{N}\left[\frac{1}{d_{l}^{2}\left(d_{l}^{2}-1\right)}\tr\left\{ \rho P_{l}\right\} ^{2}\tr\left\{ BP_{l}\right\} ^{2}\right.\nonumber \\
 & -\frac{1}{d_{l}\left(d_{l}^{2}-1\right)}\tr\left\{ \rho P_{l}\right\} ^{2}\tr\left\{ BP_{l}BP_{l}\right\} \nonumber \\
 & -\frac{1}{d_{l}\left(d_{l}^{2}-1\right)}\sum_{a,b}\left|\left\langle l,a\right|\rho\left|l,b\right\rangle \right|^{2}\tr\left\{ BP_{l}\right\} ^{2}\nonumber \\
 & +\left.\frac{1}{d_{l}^{2}-1}\sum_{a,b}\left|\left\langle l,a\right|\rho\left|l,b\right\rangle \right|^{2}\tr\left\{ BP_{l}BP_{l}\right\} \right]\,.\label{eq:43}
\end{align}
The very first summand of the above expression, $\left[S_{2}\right]_{U}$,
can be used, combined with the first case, $\left[S_{1}\right]_{U}$,
to eliminate the second part of the variance, $\mu_{t}^{2}$:
\begin{align}
\sigma_{t}^{2}=\left|\left[I_{1}\right]_{U}+\left[I_{2}\right]_{U}-\mu_{t}^{2}+\left[I_{3}\right]_{U}\right| & \leq\left|S_{2}^{'}\right|+\left|\left[I_{3}\right]_{U}\right|\,.\label{eq:24}
\end{align}

To start estimating an upper bound we notice that there exists a self-adjoint
and non-negative operator $\rho^{1/2}$ which fulfills $\rho^{1/2}\rho^{1/2}=\rho$.
This enables us to use the Cauchy-Schwarz inequality on 
\begin{equation}
\sum_{a,b}\left|\left\langle l,a\right|\rho\left|l,b\right\rangle \right|^{2}\leq\sum_{a,b}\left\langle l,a\right|\rho\left|l,a\right\rangle \left\langle l,b\right|\rho\left|l,b\right\rangle =\tr\left\{ \rho P_{l}\right\} ^{2}.\label{eq:25}
\end{equation}
Also, we state that the variance is invariant under a shift $B\rightarrow B+1_{\mathcal{H}}c$,
where $c$ is an arbitrary real number. We perform this shift such
that $B$ will be positive definite. Then we observe that 
\begin{equation}
\tr\left\{ CD\right\} \leq\left|D\right|\tr\left\{ C\right\} ,\label{eq:26}
\end{equation}
with $\left|D\right|$ denoting the operator norm and with $C$ as
well as $D$ being self-adjoint and non-negative operators. One readily
verifies that $P_{l}$ and $BP_{l}B$ are self adjoint and non-negative
as well. Applying all this we can bound 
\begin{align}
\left|S_{2}^{'}\right|\leq\sum_{l=1}^{N}\left(\frac{p_{l}^{2}\left|B\right|^{2}d_{l}^{2}}{d_{l}^{2}\left(d_{l}^{2}-1\right)}+\frac{p_{l}^{2}\left|B\right|^{2}\left(d_{l}+d_{l}^{2}\right)}{d_{l}\left(d_{l}^{2}-1\right)}+\frac{p_{l}^{2}\left|B\right|^{2}d_{l}}{d_{l}^{2}-1}\right)
\end{align}
For $d\geq2$ and the correct choice of $c$, such that $\left|B\right|=\db$,
we find 
\begin{equation}
\left|S_{2}^{'}\right|\leq4\left(\db\right)^{2}\underset{n}{\max}\left(\frac{p_{n}}{d_{n}}\right).\label{eq:28}
\end{equation}

This leaves us to evaluate the third case, namely $S_{3}$, 
\begin{align}
\left[S_{3}\right]_{U} & =\sum_{l\neq j}\sum_{\alpha,\beta,a,b}\left\langle l,a\right|\rho\left|j,b\right\rangle \left\langle j,\beta\right|B\left|l,\alpha\right\rangle \sum_{\gamma,\omega,c,d}\left\langle j,c\right|\rho\left|l,d\right\rangle \left\langle l,\omega\right|B\left|j,\gamma\right\rangle \cdot\left[U_{j,b\beta}U_{l,d\omega}U_{l,a\alpha}^{*}U_{j,c\gamma}^{*}\right]_{U}
\end{align}
The unitary average here amounts to the same calculation as in equations
(\ref{eq:18}) and (\ref{eq:21}): 
\begin{align}
\left[U_{j,b\beta}U_{l,d\omega}U_{l,a\alpha}^{*}U_{j,c\gamma}^{*}\right]_{U} & =\left[U_{j,b\beta}U_{j,c\gamma}^{*}\right]_{U_{j}}\left[U_{l,d\omega}U_{l,a\alpha}^{*}\right]_{U_{l}}=\frac{1}{d_{j}}\delta_{bc}\delta_{\beta\gamma}\frac{1}{d_{l}}\delta_{da}\delta_{\omega\alpha}\,,
\end{align}
implying that 

\begin{align}
\left|\left[S_{3}\right]_{U}\right| & =\left|\sum_{l\neq j}\sum_{\alpha,\beta,a,b}\frac{1}{d_{l}d_{j}}\left\langle l,a\right|\rho\left|j,b\right\rangle \left\langle j,\beta\right|B\left|l,\alpha\right\rangle \left\langle j,b\right|\rho\left|l,a\right\rangle \left\langle l,\alpha\right|B\left|j,\beta\right\rangle \right|
\end{align}
Now we invoke the Cauchy-Schwarz inequality from (\ref{eq:25}) again
to arrive at 
\begin{align}
\left|\left[S_{3}\right]_{U}\right| & \leq\sum_{l=j}\frac{1}{d_{l}d_{j}}\tr\left\{ \rho P_{l}\right\} \tr\left\{ \rho P_{j}\right\} \left|\tr\left\{ BP_{l}BP_{j}\right\} \right|\\
 & \leq\underset{n}{\max}\left(\frac{p_{n}}{d_{n}}\right)\sum_{j}\frac{1}{d_{j}}\tr\left\{ \rho P_{j}\right\} \tr\left\{ B\sum_{l}P_{l}BP_{j}\right\} \label{eq:34}\\
 & \leq\underset{n}{\max}\left(\frac{p_{n}}{d_{n}}\right)\left(\db\right)^{2}.\label{eq:35}
\end{align}
In (\ref{eq:34}) we noticed that $\tr\left\{ BP_{l}BP_{j}\right\} $
has to be non-negative since $BP_{l}B$ as well as $P_{j}$ are non-negative
operators. To achieve equation (\ref{eq:35}) we made use of (\ref{eq:26}),
$\left|B^{2}\right|=\left|B\right|^{2}$ and $\left|B\right|=\db$. 

Recalling (\ref{eq:24}) and making use of (\ref{eq:28}) as well
as (\ref{eq:35}) we obtain the following bound for the variance 
\begin{equation}
\sigma_{t}^{2}\leq5\left(\db\right)^{2}\underset{n}{\max}\left(\frac{p_{n}}{d_{n}}\right).
\end{equation}
Finally, using \eqref{21} we reach \eqref{9}, \eqref{10}.

\bigskip

\vspace*{1cm}
ACKNOWLEDGMENTS:
This work was supported by the 
Deutsche Forschungsgemeinschaft (DFG)
under Grants No.\ RE 1344/10-1,
RE 1344/12-1, GE 1657/3-1, STE 2243/3-1 
within the Research Unit FOR 2692, 
and by the Studienstiftung 
des Deutschen Volkes.


\end{document}